\documentclass[11pt]{article}
\usepackage{psfig}
\usepackage{amsmath, amssymb}
\begin{document}
\title{X-ray Flux Related Timing and Spectral Features of 2S 1417-62}
\author{\.{I}nam, S. \,{C}a\u{g}da\c{s} $^1$; Baykal, Altan $^1$; Scott, 
 D. Matthew $^2$ 
; Finger, Mark $^2$; Swank, Jean $^3$ \\ \small{$^1$ Department of Physics, Middle East Technical University, Inonu
Bulvari, Balgat, Ankara 06531 Turkey} \\
\small{$^2$ National Space Science and Technology Center, 320 Sparkman Dr.,
Huntsville AL 35805 USA} \\ \small{$^3$ NASA Goddard 
Space Flight Center, Greenbelt, MD 20771 USA}}
\date{9 October 2003} 
\maketitle
\abstract{RXTE observations of the X-ray transient pulsar 2S 
1417-62 between 1999 November and 2000 
August with a total exposure of $\sim 394$ ksec were analyzed. Observations 
include a main outburst followed by a series of mini outbursts. Changes in 
pulse morphology and pulse fraction were found 
to be related to the changes in X-ray flux. Particularly low X-ray 
flux regions were found to have significantly lower pulse fractions with 
different pulse morphologies.  The 3-60 keV PCA-HEXTE  main outburst spectrum was modeled 
with an absorbed power law model with high energy cut-off and a Gaussian 
Iron line complex feature. Using the same spectral model, individual 3-20 keV PCA 
spectra were 
found to be softer and less absorbed in low X-ray flux regions between 
outbursts. Spectral studies showed that hydrogen column density was correlated,
and the power law index was anti-correlated with the 3-20 keV X-ray flux.
X-ray flux related spectral and timing features in 2S 1417-62 except for low 
X-ray flux regions were interpreted as a sign of disc accretion with 
a similar accretion geometry with a varying mass accretion rate ($\dot{M}$),
whereas spectral and timing features of the low X-ray flux regions were 
interpreted as a sign of possible temporary accretion geometry change prior 
to the next 
periastron where $\dot{M}$ increases again to restore the original accretion 
geometry. }

\section{Introduction}
The X-ray source 2S 1417-62 was detected by SAS-3 in 1978 (Apparao et al. 
1980). Analysis of the SAS 3 observations showed evidence of $\sim 57$ mHz 
pulsations (Kelley et al. 1981). Einstein and optical observations 
identified a Be star companion at a distance of 1.4-11.1 kpc (Grindlay et 
al. 1984). From the timing analysis of BATSE observations between August 
26, 1994 and July 7, 1995, orbital parameters were determined and a 
correlation was found between spin-up rate and pulsed flux (Finger, 
Wilson \& Chakrabarty 1996). Orbital period and eccentricity of the source 
were found to be 42.12 days and 0.446 respectively. 

A "Be star" is an early type non-supergiant star which has a 
circumstellar disc emanating from its rotational equator which is thought to be 
formed possibly by fast spin rotation, non-radial pulsations or magnetic loops 
(Slettebak 1988). Most of the Be/X-ray binary pulsar systems like 2S 1417-62 
show recurrent X-ray outbursts that are thought to be mainly due to the 
fact that the neutron star is accreting material from the Be 
star's circumstellar disc (Negueruela 1998).

In this work, we will present and discuss our timing and spectral 
analysis of RXTE PCA observations of 2S 1417-62. In Section 2, we will 
present our timing and spectral analysis. In Section 3, we will discuss our 
results, and derive some conclusions.

\begin{figure}[h]
\psfig{file=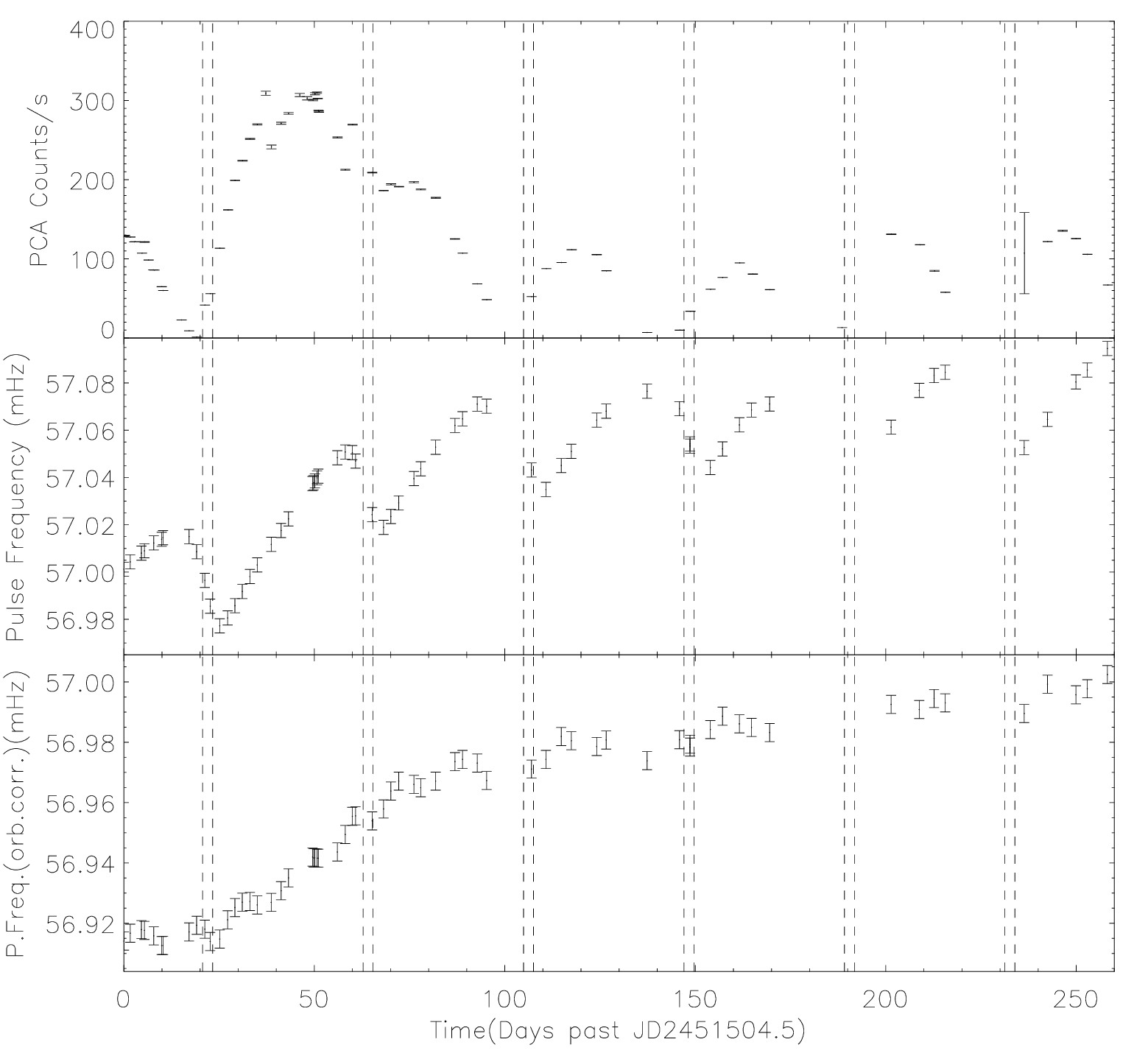,height=14cm,width=12cm} 
Fig. 1.--- Background subtracted PCA count rate normalized to 5 PCUs in 3-20 keV band, 
frequency history before and after orbit subtraction are plotted against 
time. 
Initial time value corresponds to November 22, 1999, while the peak 
of the main outburst is about 50 days later than this date. Note that in 
some parts of the data, frequencies could not be estimated. These 
frequency values were left blank. All of the errors are in $1\sigma$ level.
Vertical dashed lines indicate the orbital phase corresponding to periastron
passages. 
\end{figure}
\section{Observations and Data Analysis}
We analyzed RXTE archival observations of 2S 1417-62 between November 
22,1999 and August 7, 2000 (See Table 1 for observation list). The results 
presented here are based on data collected
with the Proportional Counter Array (PCA; Jahoda et al., 1996) and the High Energy X-ray
Timing Experiment (HEXTE; Rotschild et al. 1998). The PCA 
instrument consists of an array 
of 5 proportional counters (PCU) operating in the 2-60 keV energy range, with 
a total effective area of approximately 7000 cm$^{2}$ and a field of view 
of $\sim 1^{\circ}$ FWHM. Although the number of active PCU's 
varied between 1 and 5 during the observations, observations after 
13 May 2000 belongs to the observational epoch for which background level
for one of the PCUs (PCU0) increased due to the fact that this PCU started
to operate without a propane layer. Latest combined background models
(CM) were used
together with the latest FTOOLS release (5.2) to estimate the 
appropriate background. 
The HEXTE instrument consists of two independent
clusters of detectors, each cluster containing four NaI(Tl)/CsI(Na) phoswich
scintillation counters (one of the detectors in cluster 2 is not used for
spectral information) sharing a common $\sim 1^{\circ}$ FWHM. The field of
view of each cluster was switched on and off source to provide background 
measurements. The net open area of the seven detectors used for spectroscopy
is 1400 cm$^2$. Each detector covers the energy range 15-250 keV.

Analyzed observations consist of a main outburst that lasted for two 
orbital periods, followed by a series of mini outbursts each lasting for 
a single orbit. Duration of the main outburst and the mini outbursts are 
consistent with those covered in 1994 BATSE observations.

\subsection{Timing Analysis}

\begin{table}
\caption{Observation List for 2S 1417-62}
\label{Pri}  
\[
\small{
\begin{tabular}{c c c}\hline
Time of Observation & Exposure & XTE Obs ID \\
 day/month/year      &  ksec  &     \\ \hline \hline
01-29/12/1999 & 58.7 & 40051  \\
31/12/1999-19/01/2000 & 83.2 & 40070 \\
22-30/11/1999 \& & 125.0 & 40436 \\
25/01-29/02/2000 & & \\
03/03-07/08/2000 & 127.1 & 50095 \\
\hline
\end{tabular}}
\]
\vspace{-0cm}
\end{table}

Background light curves were generated using
the background estimator models based on the rate of very large events,
spacecraft activation and cosmic X-ray emission with the standard PCA
analysis tools and were subtracted from the source light curve obtained 
from
the event data. The background subtracted light
curves were corrected to the barycenter of the solar system.

Pulse frequencies for 2S 1417-62 were found by folding the time series 
on statistically independent trial periods (Leahy et al. 1983).
Master pulses were constructed from these observations
by folding the data on the period giving the maximum $\chi^2$.
The master pulses were arranged in 20 phase
bins, represented by their Fourier harmonics (Deeter \& Boynton 1985), and
cross-correlated with the harmonic representation of average pulse 
profiles
from each observation. The pulse arrival times were obtained from the 
cross-correlation analysis.The observations of 2S 1417-62 are irregularly spaced, with many gaps of a week or more. In these gaps, pulse frequency derivatives can cause 
cycle count ambiguity. Therefore, we obtained the pulse frequencies using 
the pulse arrival times from short segments of the data spans which are 
typically a couple of days. Then the resulting pulse frequencies were 
orbitally corrected using the binary orbit parameters found from BATSE 
observations. In fitting of pulse frequencies for the orbit model, 
we left the orbital epoch as free parameter and used the the orbital 
parameters given by Finger et al. (1996). We found the new epoch to be JD $2551612.67\pm 0.05$ and the new orbital period to be $42.19\pm 0.01$ days, while other orbital parameters were left unchanged. Evolution of PCA count rate, pulse frequency prior to orbital correction, and orbitally corrected pulse 
frequency were plotted in Figure 1.

\begin{figure}[h]
\begin{tabular}{lll}
\psfig{file=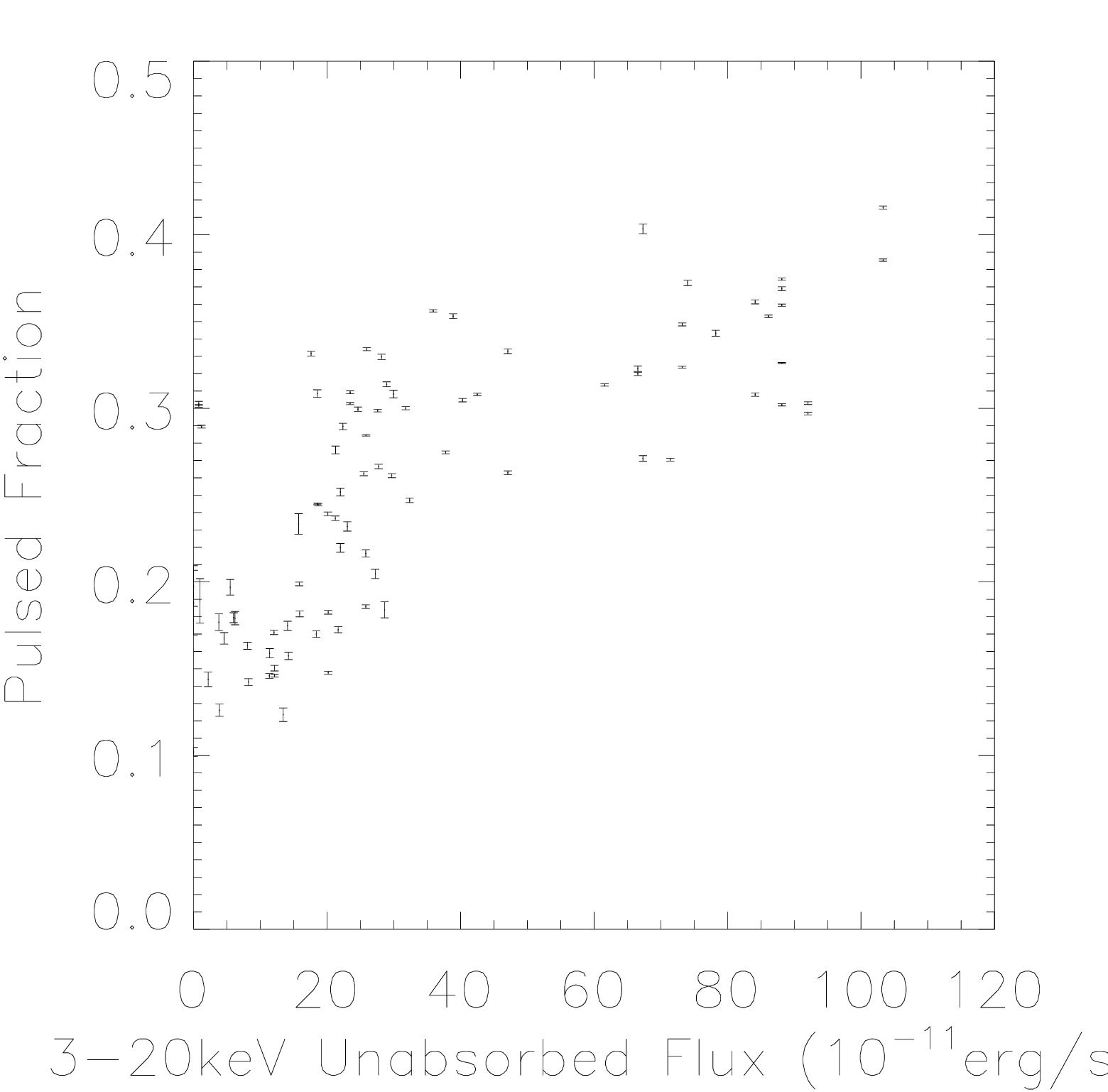,height=6.5cm,width=5.5cm} &
\psfig{file=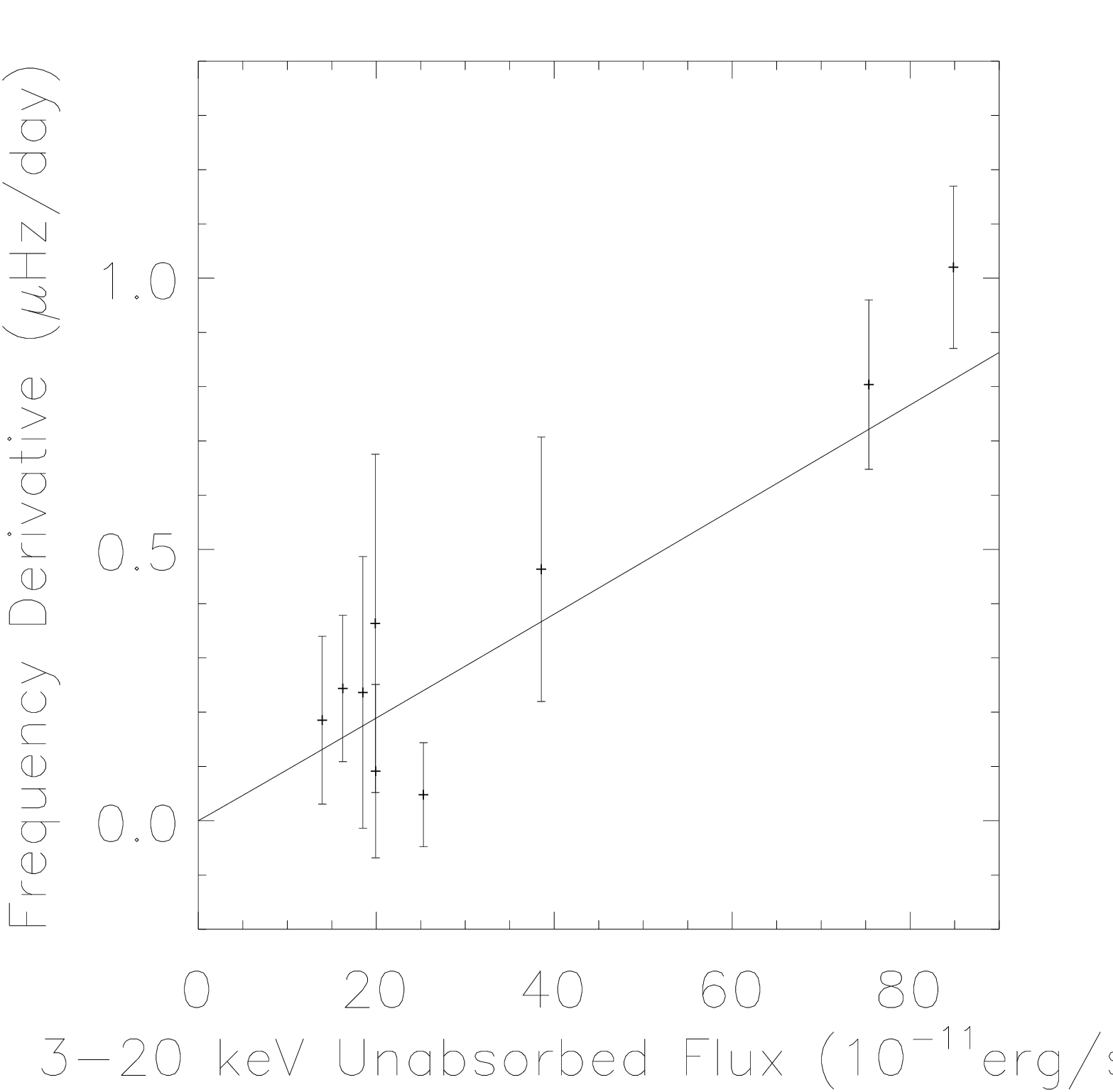,height=6.5cm,width=5.5cm} &
\psfig{file=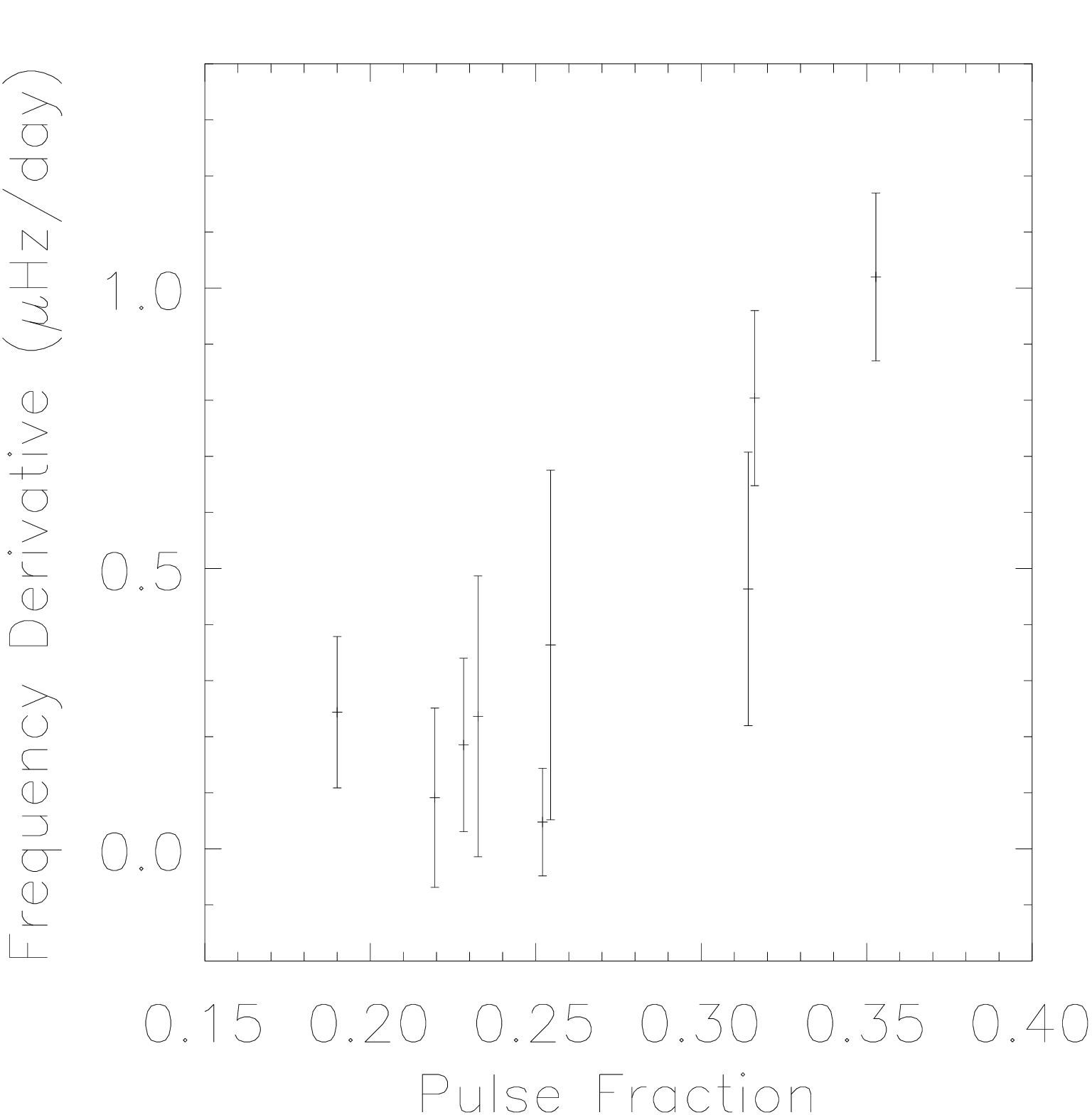,height=6.5cm,width=5.5cm} \\
\vspace{0.3cm}
\end{tabular}
Fig. 2 -- {\bf{(left)}} Pulse fraction versus 3-20 keV unabsorbed X-ray 
flux. $(F_{max}-F_{min})/(F_{max}+F_{min})$ definition is used to calculate 
pulse fractions where $F_{max}$ and $F_{min}$ are the highest and lowest 
fluxes of the phase bins.{\bf{(middle)}} Pulse Frequency Derivative 
versus 3-20 keV unabsorbed X-ray flux. Frequency derivatives were found by linear 
fitting of $\sim 20$ day intervals of frequency history. X-ray fluxes were 
found by averaging the corresponding X-ray flux values. Solid line indicates
best power law fit with the power index 1.01. {\bf{(right)}} Pulse frequency 
derivative versus 
pulse fraction. Pulse frequency derivatives from the middle panel of this 
figure were plotted against pulse fractions which were found by averaging 
corresponding pulse fraction values. Errors in all of the panels are in $1\sigma$ level.
\end{figure}

\begin{figure}[h]
\begin{tabular}{cc}
\psfig{file=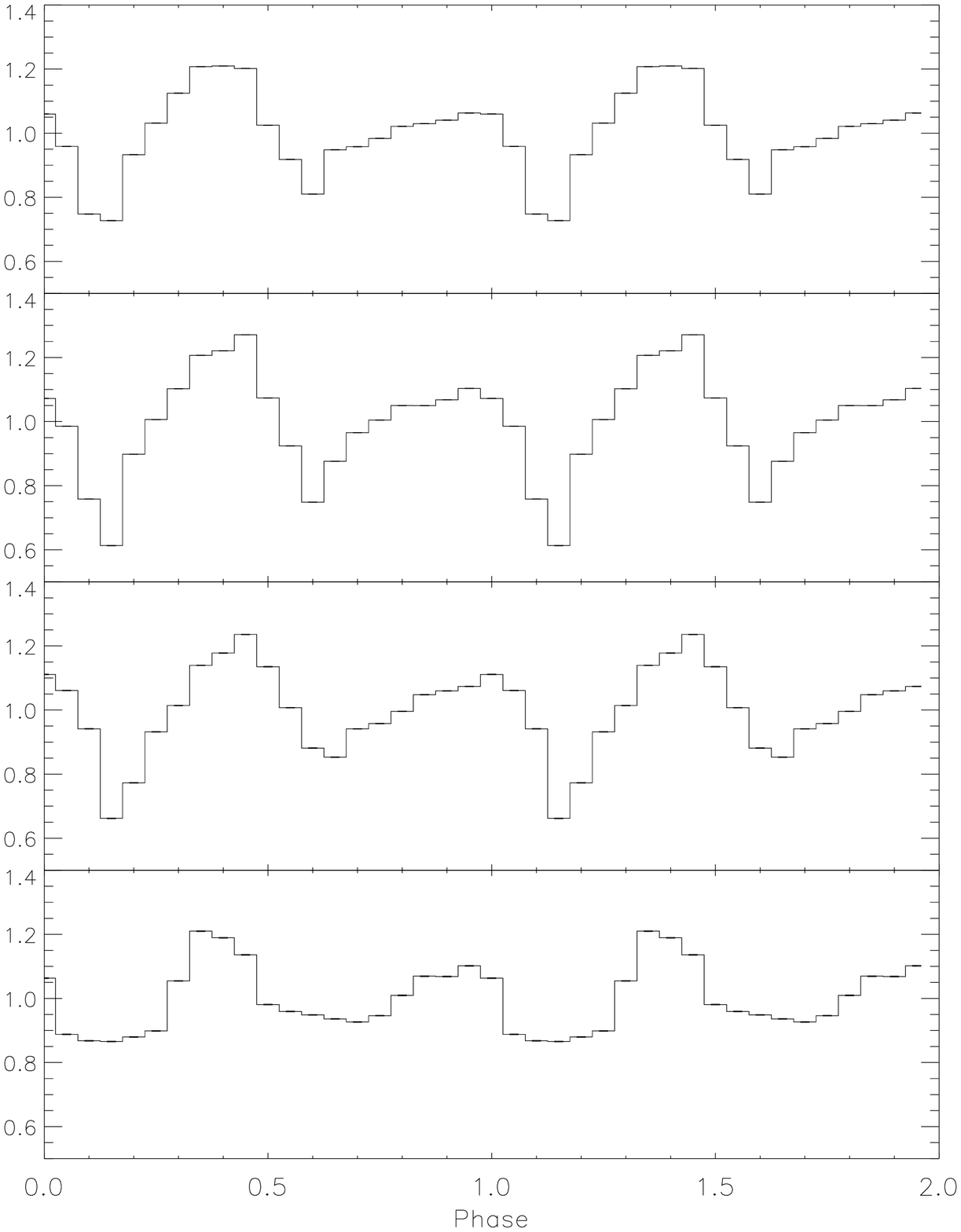,height=14cm,width=9cm} &
\psfig{file=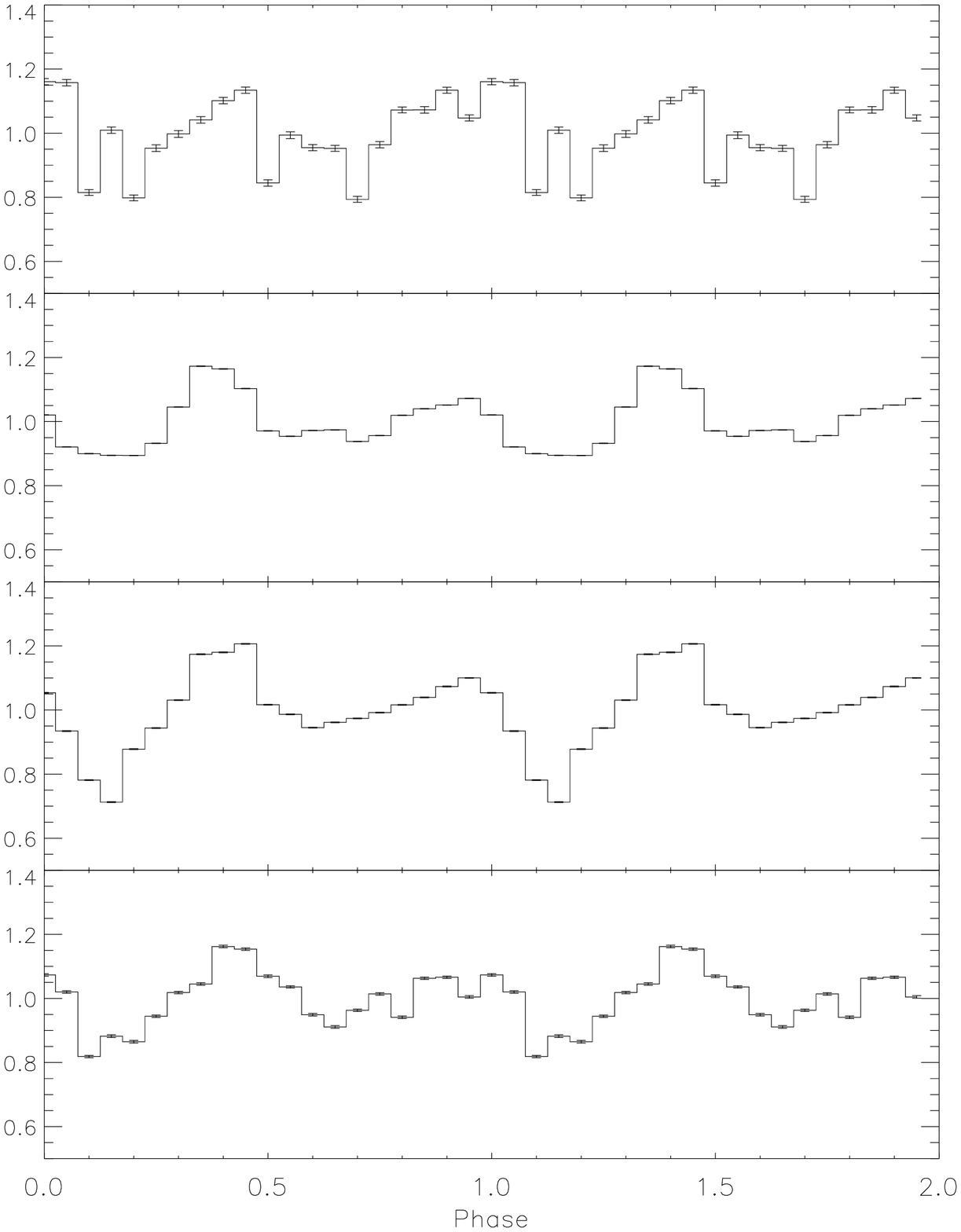,height=14cm,width=9cm} 
\end{tabular} \\
Fig. 3 -- {\bf{(left)}} Pulse profiles for days $\sim 31.2$ 
(rise of the mainoutburst), 
$\sim 50.6$ (peak of the main outburst), $\sim 76.2$ (peak of the main outburst 
at the second orbit), $\sim 95.3$ (decline of the main outburst), {\bf{(right)}} 
$\sim 101.9$ (low X-ray flux part), $\sim 107.1$ (rise of the mini outburst),
$\sim 124.1$ (peak of the mini outburst), and $\sim 137.4$ (decline of the mini 
outburst). Corresponding 3-20 keV unabsorbed fluxes in units $10^{-11}$ergs.cm$^2$.s$^{-1}$ are 75.1, 105.1, 54.0, 11.9, 1.18, 11.9, 21.9, and 3.69 respectively. Phases
of the pulse profiles were manually aligned, using the similarities in the
pulse shapes and the positions of primary and secondary peaks.
\end{figure}

Individual pulse profiles were used to calculate pulse fractions. The relation 
between pulse fraction and 3-20 keV unabsorbed X-ray flux was plotted in the left panel 
of Figure 2. 8 pulse profiles corresponding to different flux levels were 
plotted in Figure 3. 
Spin-up rates were found by linear fitting of the 
$\sim 20$ day long segments of 
frequency time series. In these time intervals, 3-20 keV unabsorbed X-ray flux and pulse fraction values were 
recalculated by averaging corresponding unabsorbed X-ray flux and pulse fraction values. 
Relations among frequency derivative, recalculated pulse 
fraction and recalculated 3-20 keV unabsorbed X-ray flux were plotted in the middle and 
right panels of Figure 2. 

\subsection{Spectral Analysis}
Spectrum, background and response matrix files were created using 
{\it{FTOOLS}} 5.2 data analysis software. Background spectra were 
generated using the background estimator models based on the rate of very 
large events, spacecraft activation and cosmic X-ray emission. 

\begin{table}
\caption{Spectral Parameters of PCA Observations of 2S 1417-62}
\label{Pri}  
\[
\scriptsize{
\begin{tabular}{c c c c c c c}\hline \hline
Time & $n_H$  & Fe Energy  & Fe Sigma & Fe Norm. 
& PL Index  & \\
(day) & ($10^{22}$cm$^{-2}$) & (keV) & (keV) & (cts.cm$^{-2}$.s$^{-1}$) &  &\\ \hline
50.7 & $4.22\pm 0.23$ & $6.75\pm 0.12$ & $1.15\pm 0.13$ & $(1.38\pm 
0.41)10^{-3}$ & $0.90\pm 0.02$ & \\
101.9 & $0.24\pm 0.24$ & $6.48\pm 0.25$ & $0.22\pm 0.20$ & $(4.98\pm  
2.50)10^{-5}$ & $2.00\pm 0.30$ & \\
126.7 & $4.53\pm 0.61$ & $6.87\pm 0.10$ & $0.96\pm 0.28$ & $(3.18\pm 
1.50)10^{-4}$ & $1.27\pm 0.05$ & \\
\hline \hline
Time & PL Norm  & $E_{{\mathrm{cut}}}$  & $E_{{\mathrm{fold}}}$  & Reduced $\chi^2$ 
&  3-20 keV Flux & 3-20 keV Flux \\
 & & & & & (absorbed) & (unabsorbed) \\ 
(day) & (cts.keV$^{-1}$.cm$^{-2}$.s$^{-1}$) & (keV) & (keV) & (30 d.o.f) & 
(ergs.cm$^{-2}$.s$^{-1}$) & (ergs.cm$^{-2}$.s$^{-1}$) 
\\ \hline
50.7 &  $(3.31\pm 0.15)10^{-2}$ & $11.02\pm 0.13$ & $22.71\pm 0.63$ & 1.56 & 
$9.79\times 10^{-10}$ & $1.05\times 10^{-9}$\\
101.9 & $(3.72\pm 1.12)10^{-3}$ & $18.03\pm 5.00$ & $30.40\pm 6.00$ & 0.68 & 
$1.17 \times 10^{-11}$ & $1.17 \times 10^{-11}$ \\
126.7 &  $(1.57\pm 0.18)10^{-2}$ & $10.97\pm 0.72$ & $22.27\pm 1.95$ & 1.05 
& $1.98 \times 10^{-10}$ & $2.12 \times 10^{-10}$ \\
\hline \hline
\end{tabular}}
\]
\vspace{-0cm}
\end{table}

Spectral analysis and error estimation of the spectral parameters were 
performed using {\it{XSPEC}} version 11.1. Overall PCA (3-35 keV) and HEXTE
(25-60 keV) spectra were constructed for the main outburst covered by the dataset 40070 
(Figure 4).  Individual PCA spectra were constructed 
for the same time intervals that were used for calculating arrival times.  
Energy channels corresponding to 3-20 keV energy range were used to fit 
the individual PCA spectra. Energies lower than 3 keV were ignored due to uncertainties 
in background modelling while energies higher than 20 keV were ignored as 
a result of poor counting statistics. No systematic error was added to 
the errors.

A power law model with low energy absorption (Morrison \& McCammon, 1983), multiplied by 
an exponential high energy cut-off function (White et al. 1983) was used
to model both the overall 3-55 keV (PCA-HEXTE) spectrum and
the individual 3-20 keV band (PCA) spectra. Additional 
Iron emission line complex modeled as a Gaussian at $\sim 6.4-6.8$ keV was 
required as the part of the spectral model. Evolution of spectral parameters of 
3-20 keV PCA spectrum 
was plotted in Figure 5. Spectral parameters corresponding to the overall
PCA-HEXTE spectra and these 3 regions were listed in Table 2 and 3 
respectively. Using individual 3-20 keV spectra, the hydrogen column density was 
found to be correlated with the 3-20 keV X-ray flux, while the power-law index 
was found to be anti-correlated  with the 3-20 keV X-ray flux (see Figure 6). 

\begin{table}
\caption{Spectral Parameters of PCA-HEXTE Observations of the main outburst of 
2S 1417-62}
\label{Pri}  
\[
\footnotesize{
\begin{tabular}{l c}\hline \hline
Parameter & Value \\ \hline
Hydrogen Column Density ($10^{22}$cm$^{-2}$) & $3.99\pm 0.14$ \\
Gaussian Line Energy (keV) & $6.82\pm 0.16$ \\
Gaussian Line Sigma (keV) & $1.08\pm 0.15$ \\
Gaussian Normalization (cts.cm$^{-2}$.s$^{-1}$) & $(1.20\pm 0.15)\times 10^{-3}$
\\
Power Law Photon Index & $0.90\pm 0.02$ \\
Cut-off Energy (keV) & $11.1\pm0.6$ \\
E-folding Energy (keV) & $21.6\pm1.3$ \\
Power Law Normalizations (cts.keV$^{-1}$.cm$^{-2}$.s$^{-1}$) & $(2.77\pm 0.23)\times 10^{-2}$ \\
Calculated X-ray Fluxes (3-60 keV in ergs.cm$^{-2}$.s$^{-1}$) & \\
--- Absorbed & $1.59\times 10^{-9}$ \\
--- Unabsorbed & $1.65\times 10^{-9}$ \\
Reduced $\chi^2$ & 1.01 (119 d.o.f) \\ \hline \hline
\end{tabular}}
\]
\vspace{-0cm}
\end{table}

Using XSPEC software and assuming a distance of 11 kpc,  3-20 keV band 
luminosities were 
found to be $\simeq 1.4\times 10^{34}$ ergs/s for the low X-ray flux regions 
between outbursts, $\simeq 
1.3\times 10^{36}$ ergs/s for the peak of the main outburst, and 
$\simeq 2.3\times 10^{35}$ ergs/s for the peak of the first mini 
outburst followed by the main outburst.

\begin{figure}
\begin{center}
\psfig{file=fig4_revised.eps,height=8cm,width=12cm,angle=0,angle=-90}
\end{center}
Fig. 4 -- Combined PCA and HEXTE spectrum for the dataset 40070. The 
bottom panel shows the residuals of the fit in terms of $\sigma$ values.
\end{figure}

\begin{figure}
\begin{center}
\psfig{file=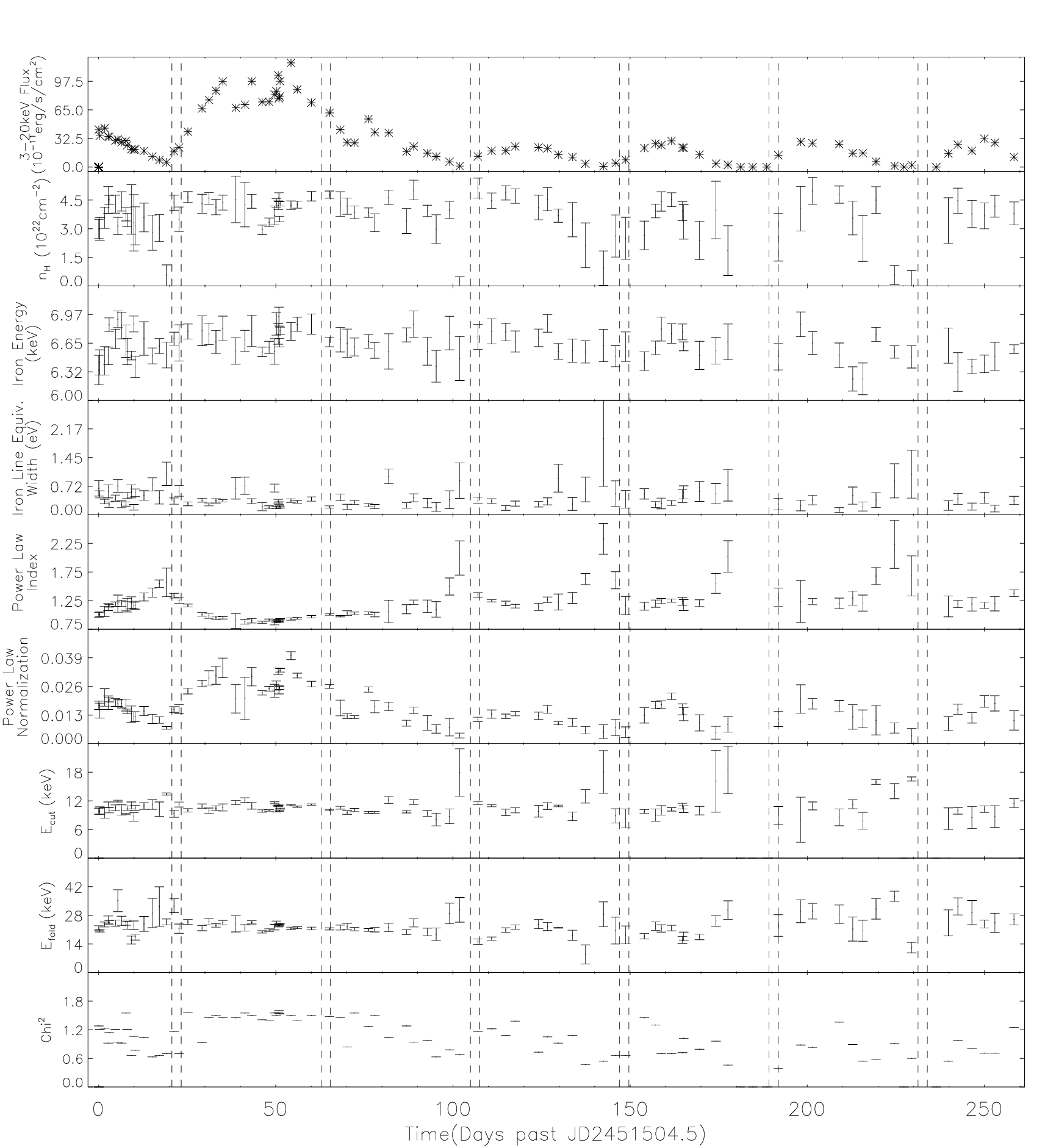,height=12.5cm,width=14.5cm,angle=0}
\end{center}
Fig. 5 -- Evolution of Hydrogen column density, Iron line complex peak 
energy, iron line equivalent width, power law index, power law 
normalization, cut-off energy, folding energy, and reduced $\chi^2$. Vertical
dashed lines indicate the orbital phase corresponding to periastron passages. 
\end{figure}

\begin{figure}
\begin{center}
\begin{tabular}{cc}
\psfig{file=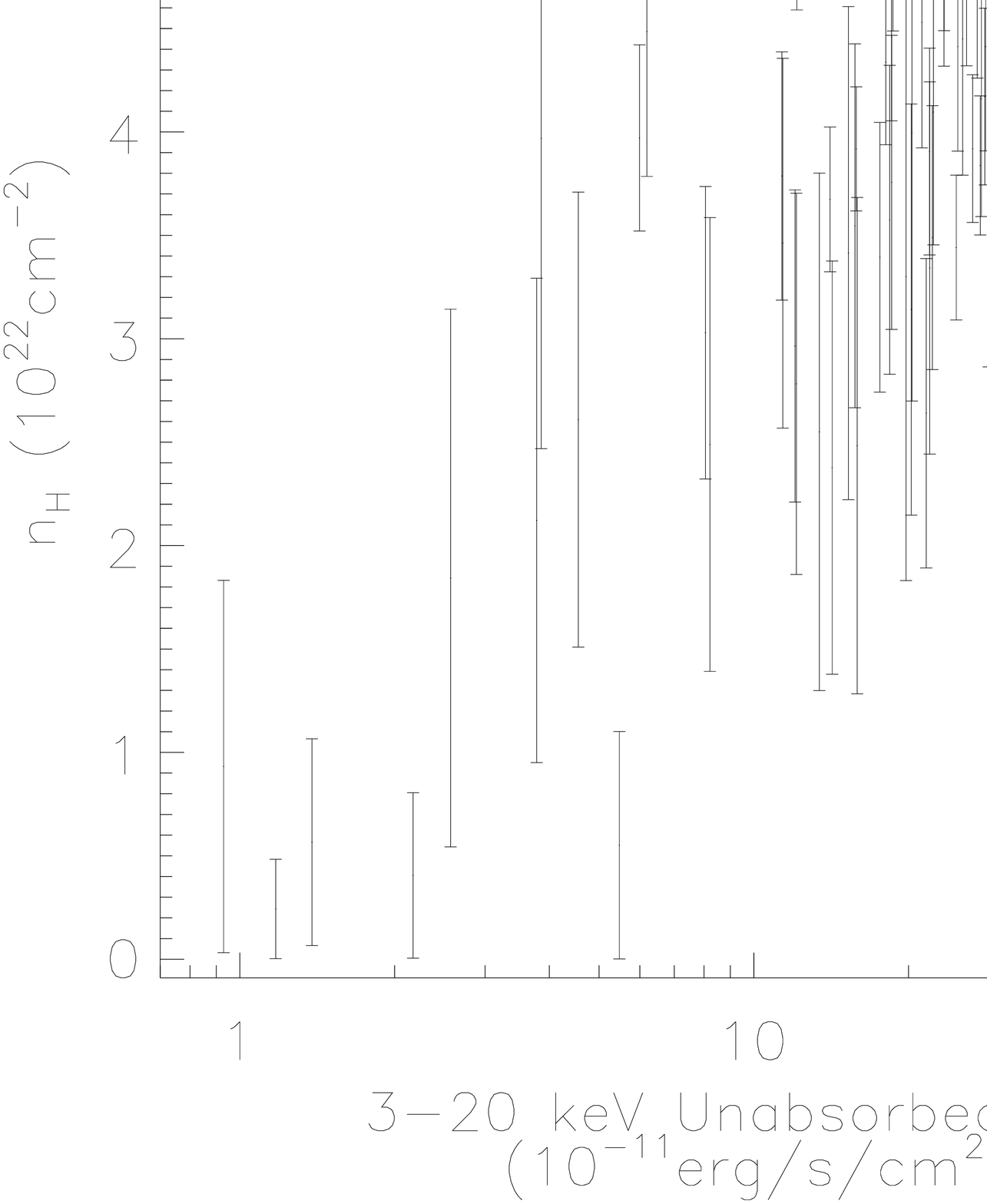,height=6cm,width=7cm,angle=0} & \hspace{0.5cm}
\psfig{file=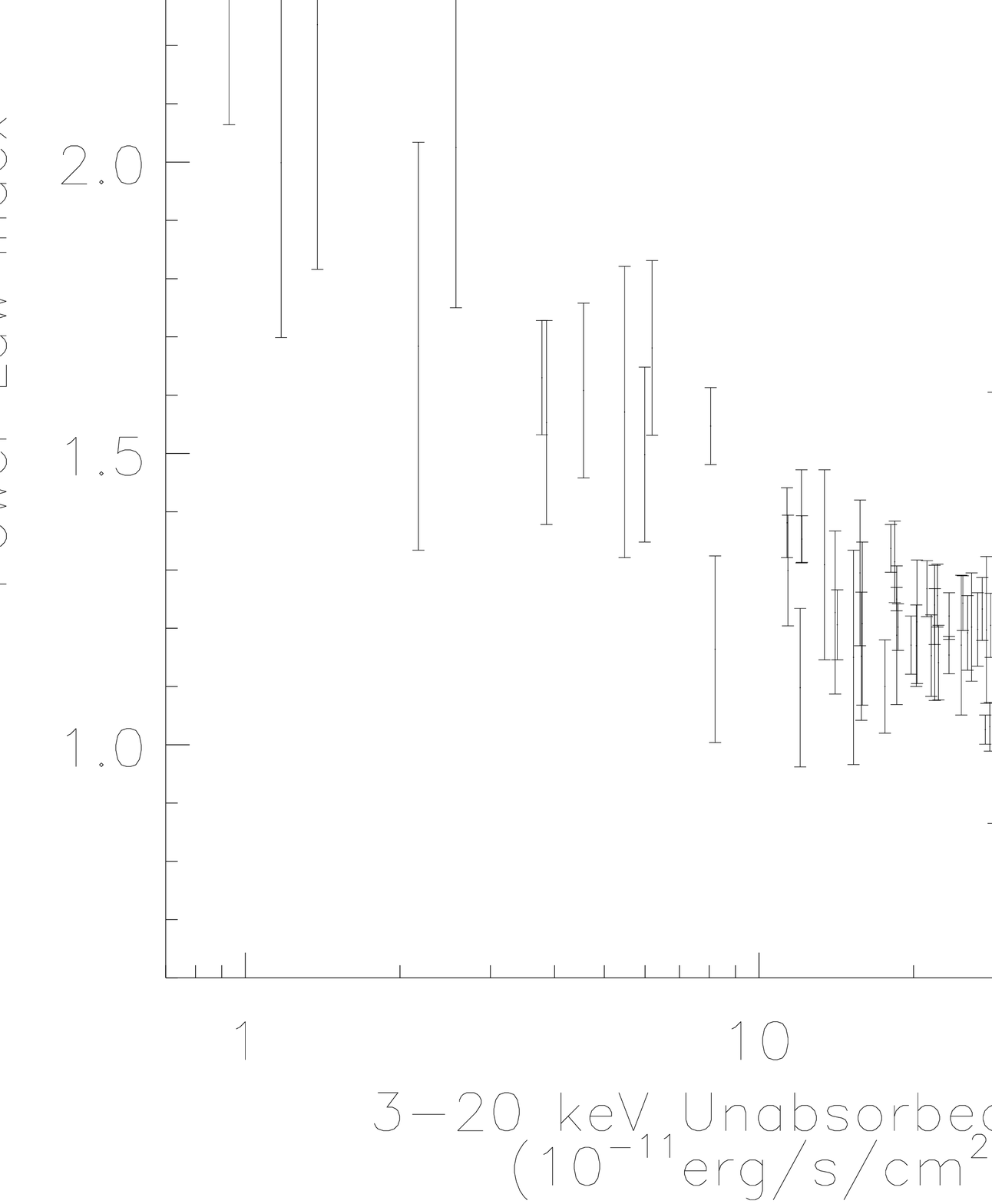,height=6cm,width=7cm,angle=0}
\end{tabular} \\
\end{center}
Fig. 6 -- {\bf{(left)}} Hydrogen column density as a function of 3-20 keV 
unabsorbed X-ray flux. {\bf{(right)}} Power law index as a function of 3-20 keV unabsorbed X-ray flux.
\end{figure}

\section{Discussion and Conclusion}
In a Be/neutron star binary system, occasional
X-ray outbursts may be observed, implying the neutron
star is accreting material, presumeably from the Be star's circumstellar 
disc if the circumstellar disc undergoes a sufficiently large increase in 
its radial extent and density to intersect the neutron stars orbital path. 
The neutron
star may stop accreting and ceases to be apparent as a bright X-ray source
after the Be circumstellar disc has contracted. 

The pattern of X-ray outbursts is affected by the size, eccentricity and 
orientation of the neutron star's orbit with respect to the Be star. The
orbit could be coplanar with the Be circumstellar disc or offset such
that the neutron star may pass through the circumstellar disc. 
In the latter case, neutron star may either pass through the disc twice per orbit or 
pass only once per orbit as in the case for 2S 1417-62.

There are basically two accretion modes possible for the neutron star in a Be/Neutron 
star binary system. The first possible mode is direct 
accretion of the Be circumstellar disc material by the neutron star during the 
circumstellar disc passage in a manner similar to a wind-fed pulsar experiencing
Bondi-Hoyle type accretion. For this mode, there will be less efficient 
angular momentum transfer compared to a standard disc accretion. This mode
accretion may be considered to be the cases for the X-ray spikes during the periastron 
and for the mini outbursts of 2S 1845-024 (Finger et al. 1999).

The second mode that actually seems to be the case for 2S 1417-62 is accretion 
after the 
bound material has collapsed into a standard but temporary accretion disc. An accretion 
disc can transfer 
angular momentum either at the vicinity of the magnetospheric radius where 
the disc began to be disrupted and the material is channeled from the inner 
edge of the disc to the magnetic poles (via "material torque"), or from the overall 
interaction of the accretion disc and magnetic field lines of the neutron star 
(via "magnetic torque"). Material torque from a prograde accretion disc, being 
proportional to mass accretion rate, always acts to spin-up the neutron star, while 
the contribution of magnetic torque from the magnetic field lines threading the 
disc outside the corotation radius $r_{co}$ is negative. Therefore, 
resultant torque can either be a spin-up (positive) or a spin-down 
(negative) torque. 

Assuming observed X-ray luminosity is proportional to the bolometric 
luminosity (or mass accretion rate), spin-up rate and X-ray flux correlation 
can be explained by accretion from accretion disks when the 
net torque is positive and of the order of the material torque (Ghosh \& Lamb 
1979; Ghosh 1993). Correlation between spin-up rate and X-ray flux in different 
X-ray energy bands has been observed in outbursts of other transient X-ray 
pulsar systems:  EXO 2030+375 (Wilson et al. 2002; Parmar,
White \& Stella 1989), A 0535+26 (Bildsten et al. 1997; Finger, Wilson \& Harmon
1996), 2S 1845-024 (Finger et al. 1999), GRO J1744-28 (Bildsten et al. 1997), 
GRO J1750-27 (Scott et al. 1997), XTE J1543-568 (In't Zand, Corbet \& 
Marshall 2001), and SAX J2103.5+4545 (Baykal, Stark \& Swank 2002). Correlation 
between 20-50 keV X-ray flux and spin-up rate was previously found for 2S 1417-62 using 
BATSE observations (Finger, 
Wilson \& Chakrabarty 1996). Our analysis of RXTE datasets of 2S 1417-62 has 
also showed the evidence of this correlation for both the main outburst and the 
following mini outbursts in 3-20 keV band (see the middle panel of Figure 2). 

When the neutron star in a Be/neutron star binary system leaves the dense 
equatorial disc of the companion, 
the accretion disc can no longer be fed by the surrounding material. In 
this case, accretion disc may disappear and
the neutron star may either continue to accrete from the non-equatorial wind of 
the companion or may 
enter the propeller phase (Illarionov \& Sunyaev 1975). In case accretion is the 
result of the companion's non-equatorial wind, it is possible to  
see erratic spin-up and spin-down episodes (just like wind accreting 
systems, see Bildsten et al. 1997; Inam \& Baykal 2000) which is not the case in 2S 
1417-62 even for the lowest flux parts of our datasets. In case wind of the companion 
does not cause accretion, propeller phase may set in. In the propeller phase, we may
expect spin pulsations to cease accompanied by a decrease in flux (Cui 1997),
although it may also be the case that the pulsations do not cease completely 
even in the propeller phase (Neguerela et al. 2000). 

Although we did not observe any erratic spin-up and spin-down episodes in 
the archival data, we found that both pulse fraction and shape of pulse 
profiles depend on the X-ray flux (Figures 2 and 3). From Figure 3, 
we found that the morphology of pulse profiles for
the peaks of the main outburst and the mini outbursts as well as the increasing 
flux regions just after the periastron and declining flux regions before the 
next periastron passages are similar, with only a 
slight difference in the pulse fractions.
However, both the pulse fraction and the morphology of the pulse profiles of the 
low flux parts of the data is
considerably different from the data elsewhere. Similarity of the pulse 
profiles of the main 
outburst and the mini outbursts may indicate that we basically have the same
accretion geometry with slightly different mass accretion rate. Together with
the fact that the spin up rate is correlated with the flux, we may argue that
we have disc accretion for all stages of the main outburst and mini outbursts
with a varying mass accretion rate but basically with the same accretion
geometry. Decreasing pulse 
fraction with changing pulse morphology for the low flux parts 
may indicate changes in accretion geometry 
and even be an indicator of a transition from disc to wind accretion, or 
an ongoing transition from accretor phase to propeller phase. SAX 
J2103.5+4545 is a counter-example for which no significant changes in pulse 
fraction and pulse profiles are seen with changing flux (Baykal, Stark \& 
Swank 2002). Therefore, accretion geometry and polar cap size of SAX 
J2103.5+4545  
should not be related considerably to the mass accretion rate while 2S 
1417-62 may have such accretion geometry changes accompanied by the decrease 
in X-ray flux. 
Correlation between spin-up rate and pulse fraction is also another evidence 
of such a transition (see right panel of Figure 2).  

Decline of the X-ray flux may be an indication of an ongoing transition from 
accretor to propeller stage. When the propeller stage sets in, the great 
majority of accreting matter cannot reach the neutron star and factor of 
decrease in the bolometric luminosity becomes 

\begin{eqnarray*}
\Delta=170M^{1/3}_{1.4}R^{-1}_6P^{2/3}_0
\end{eqnarray*}

\noindent{where $M_{1.4}$ is the mass of the neutron star in units of 
$1.4M_{\odot}$, $R_6$ is the radius of the neutron star in units of $10^6$ 
cm and $P_0$ is the spin period of the neutron star in units of second 
(Corbet 1996; Campana \& Stella 2000; Campana et al. 2002). The factor
$\Delta$ becomes $\sim 10^{3}$ for a neutron star with mass 
$\sim 1.4M_{\odot}$, radius $\sim 10^{6}$ cm and spin period of $\sim 17.5$ 
s . We do not observe such sharp X-ray flux declines in RXTE-PCA 
observations of 2S 1417-62, moreover a factor of $\sim 10^3$ decrease in 
X-ray flux of 2S 1417-62 should be considerably less than the background 
level. As a comparison, 3-20 keV X-ray flux of 2S 1417-62 is lower in the 
low X-ray flux regions between outbursts by a factor of $\sim 15$ with 
respect to peak of mini outbursts and $\sim 90$ with respect to the peak of 
the main outburst. These factors seem to be considerably smaller than the
expected factors for accretor to propeller transition. It is more likely that 
we just 
observe an accretion geometry change due to decreasing $\dot{M}$, maybe an 
ongoing transition to propeller, but not the propeller stage itself.} 

Since 2S 1417-62 is not likely to make transitions to the propeller phase, the source
is probably accreting even at the lowest flux level. In case of
accretion, the
minimum luminosities that we have should be greater than a limiting accretion
luminosity ($L_{limit}$) given as: 
(Stella et al. 1986, Illarionov, Sunyaev 1975)

\begin{equation*}
L_{min} > L_{limit} \cong 2\times 10^{37}R_6^{-1}M_{1.4}^{-2/3}P_0^{-7/3}
({\mu \over 10^{30}\rm{Gcm}^3})^{2}
\rm{ergs.s}^{-1}
\end{equation*}

\noindent{where $\mu$ ($\cong BR^3$, B being the surface magnetic
field of the neutron star) is the magnetic 
moment of the neutron star. Using the spectral model for the low flux part of the 
data and assuming a distance of 11 kpc which is within the range of possible distances 
given by Grindlay et al. (1984), 0.5-30 keV X-ray luminosity is found to
be $\sim 2.1 \times 10^{34}$ ergs/s. Using this luminosity, and assuming that 
$R\sim 10^6$cm, $M\sim 1.4M_{\odot}$,  a lower limit to the magnetic field 
can be set to be $\sim 9\times 10^{11}$G, which is consistent with the typical
magnetic fields of accretion powered pulsars.} 

From Figure 1, it is evident that the location of the periastron passage is just
after the start of both the main outburst and the mini outbursts. The mini 
outbursts last for almost an entire orbit of 2S 1417-62 and hence indicate the 
binding of 
some Be circumstellar disc material to the neutron star during disc 
passage with the accretion of the bound material occuring during the course 
of the orbit. The BATSE observations showed a very similar normal outburst 
pattern with 
the hard X-rays (above 20 keV) starting just after the periastron passage and 
ending before the next periastron passage. Pattern and duration of the main 
outburst is also consistent with the main outburst observed by BATSE, 
starting just after the periastron and lasting about two orbital periods. 
Duration of the main outburst indicate that the equatorial disc of the 
companion is dense enough to provide enough disc material initially to feed 
the disc for two orbital periods. An increase in X-ray flux during the main 
outburst is observed corresponding to the periastron passage as well, showing 
that the neutron star has entered the equatorial disc again, but, this time, 
without complete depletion of the accretion disc material around it.

Galactic coordinates (l=313.02 b=-1.598) of 2S 1417-62 reveals that the source 
is almost at the boundary of the galactic ridge region (corresponding to R1 
region defined by Valinia, Marshall 1998). Spectral model for the galactic
ridge emission includes an absorbed power law model with power law index 
$\sim 1.8$ and Raymond Smith component at $\sim 3$ keV. If the flux of the
galactic ridge were comparable to the X-ray flux in the low X-ray flux regions,
then we would not be sure about the reality of the
higher power law indices and decreasing pulse fractions, since these effects
may be due to the fact that our background is contaminated by the galactic
ridge emission. However, the X-ray flux in 3-20 keV is found to be only $\sim 2.5$\%
of the typical flux for the low X-ray flux region of 2S 1417-62 by using the
spectral model for R1 region of the galactic ridge. Therefore, it is not likely
that our results are very much affected by the galactic ridge emission.

The spectrum of 2S 1417-62 is found to be consistent with a typical accretion 
powered pulsar spectral model consisting of an absorbed power law with power 
law index $\sim 0.9-1.2$ and an high energy cut-off at $\sim 9-12$ keV. For 
the low X-ray flux regions between outbursts, the power law index is found to be 
higher reaching values of $\sim 2$, while the cut-off energy increases to $\sim 
15-18$ keV but the uncertainty in the cut-off energy increases from $\sim 0.5$ 
keV to $\sim 5$ keV meanwhile. 

Although the column density is, in general, found to be correlated with 
the X-ray flux (left panel of Figure 6), it is found to be considerably
lower for 
the low X-ray flux regions between the outbursts where it decreases to 
$\lesssim 1\times 10^{22}$ cm$^{-2}$. Lower hydrogen column densities 
together with lower X-ray flux at low X-ray flux regions may indicate lower 
matter concentration around the neutron star due to the fact that neutron 
star is in the orbital phase away from the equatorial Be wind so that 
matter around the neutron star has ceased or is about to cease. This is also
related to the fact that the X-ray flux decreases, since $\dot{M}$ of the 
accretion disc decreases due to the lower density of the surrounding material
feeding the disc.

The iron line complex at $\sim 6.4-6.8$ keV is found to exist for the overall 
data. This energy range shows that the iron line complex is composed not 
only of cold flourescent Iron K line at $\sim 6.4$ keV, but also of H-like 
and He-like Iron lines at $\sim 6.7-7.0$ keV which is possibly emitted from hot 
and ionized gas around neutron star, e.g. accretion disc corona. We are 
unable to resolve this line complex into single emission lines in each RXTE 
observation.  
Although uncertainty in iron line energy is almost comparable to the changes 
in the iron line energy, there is still a weak trend that the peak energy of 
the iron line is found to be $\sim 6.4-6.5$ keV at or 
near low X-ray flux parts of the data, while the peak energy reaches sim 
$\sim 6.7-6.8$ keV near the X-ray flux maxima of the outbursts. A plausible 
explanation for this trend of the peak energy is that density of the ionized 
gas around the neutron star increases while outbursts occur. Future observations
with new X-ray observatories like XMM-Newton and Chandra may be useful to 
resolve this iron line complex features in 2S 1417-62. 

Anti-correlation of power law index with the X-ray flux is an 
indication of the fact that the spectrum gets softened with decreasing X-ray flux. 
(right panel of Figure 6). Spectral hardening is observed 
in the low state spectra of neutron star soft X-ray transients like Aql X-1, 
which may be interpreted to be a sign of propeller stage (Zhang et al. 1998) 
or a sign of a turned-on rotation powered pulsar (Campana et al. 1998).
On the other hand, increasing power law indices (i.e softening of the spectra) 
with decreasing flux might be the consequence of mass 
accretion rate changes only (see e.g models for low luminosity X-ray sources by 
Meszaros et al. 1983; Harding et al. 1984). In this case, neither a 
transition to propeller stage nor an accretion geometry change is needed to 
explain the softening in the spectrum with decreasing flux. 
Consequently, decrease in mass accretion rate with a softening in 
the spectrum does not lead to any significant changes in pulse profiles and 
pulse fraction 
(e.g. Baykal, Stark \& Swank 2002 for SAX J2103.5+4545). We have a similar
situation in 2S 1417-62 except that in the lowest X-ray flux parts, 
the changes in pulse profiles and pulse fractions accompany spectral softening of 
2S 1417-62.  
Therefore, spectral softening observed throughout main outburst and mini 
outbursts may be similar to the case in SAX J2103.5+4545, 
however spectral 
softening in 2S 1417-62 may be the 
result of not only accretion rate changes but also accretion geometry 
changes for the low flux parts, which occur just before the periastron for 
which the neutron star 
should have accreted almost all of the accretion disc material around it.

New X-ray observations which will monitor main high outbursts, the 
following mini outbursts and X-ray dips continously so that the data will be 
continously phase-connected will be very useful for further studies of 2S 
1417-62. These observations will help us to understand the accretion 
mechanism better and test the possible transition from accretor to propeller 
stage. New observations may also be useful to revise and improve orbital 
parameters and to have a better understanding of timing and spectral 
evolution of 2S 1417-62. 

{\bf{Acknowledgments}}

S.\c{C}.\.{I}nam acknowledges the Integrated Doctorate Program scholarship from 
the Scientific and Technical Research Council of Turkey (T\"{U}B\.{I}TAK).

\section*{References}
\noindent{Apparao, K.M.V., Naranan, S., Kelley, R.L. et al., 1980, A\&A 
89,249}

\noindent{Baykal, A., Stark, M., Swank, J., 2002, ApJ 569, 903}

\noindent{Bildsten, L., Chakrabarty, D., Chiu, J. et al., 1997, ApJS 
113,367}

\noindent{Campana, S., Stella, L., Mereghetti, S., et al. 1998, ApJL 499, 65}

\noindent{Campana, L., Stella, L., 2000, ApJ 541, 849}

\noindent{Campana, S., Stella, L., Israel, G.L., et.al. 2002, ApJ 580, 389}

\noindent{Corbet, R.H.D., 1996, ApJL, 457, 31}
 
\noindent{Cui, W., 1997, ApJL, 482, 163}

\noindent{Deeter,  J.E., Boynton,  P.E., 1985, in Proc. Inuyama Workshop 
on Timing Studies of X-Ray Sources, ed. S. Hayakawa $\&$ F. Nagase 
(Nagoya: Nagoya Univ.), 29}

\noindent{Deeter, J., Crosa, L., Gerend, D. et al. 1976, ApJ, 206, 861}

\noindent{Finger, M.H., Wilson, R.B., Chakrabarty, D., 1996, A\&ASS 120, 
209}

\noindent{Finger, M.H., Wilson, R.B., Harmon, B.A., 1996, ApJ 459, 288}

\noindent{Finger, M.H., Bildsten, L., Chakrabarty, D. et al. 1999, ApJ 517,449}

\noindent{Ghosh, P., Lamb, F.K., 1979, ApJ 234, 296}

\noindent{Ghosh, P. 1993, The Evolution of X-ray binaries, eds. 
S.S.Holt \& C.S.Day, p439}

\noindent{Grindlay, J.E., Petro, L.D., McClintock, J.E., 1984, ApJ 276, 
621}

\noindent{Harding, A.K., Meszaros, P., Kirk, J.G. et al. 1984, ApJ 278, 369}

\noindent{Illarionov, A.F., Sunyaev, R.A., 1975, A\&A. 39, 185}

\noindent{\.{I}nam, S.C., Baykal, A., 2000, A\&A 353, 617}

\noindent{in't Zand, J.J.M., Corbet, R.H.D., Marshall, F.E. 2001, ApJL 
553, 165}

\noindent{Jahoda, K., Swank, J.H., Giles, A.B. et al. 1996, Proc.SPIE, 
2808, 59}

\noindent{Kelley, R.L, Apparao, K.M.V, Doxsey, R.E. et al., 1981, ApJ 
243, 251}

\noindent{Leahy, D.A., Darbo, W., Elsner, R.F. et al. 1983, ApJ 266, 160}
 
\noindent{Meszaros, P., Harding, A.K., Kirk, J.G. et al. 1983 ApjL 266, 33}

\noindent{Morrison, R., McCammon, D., 1983, ApJ 270,119}

\noindent{Negueruela, I., 1998, A\&A 338,505}

\noindent{Negueruela, I., Reig, P., Finger, M.H. et al. 2000, A\&A, 356, 1003}

\noindent{Parmar, A.N., White, N.E., Stella, L., 1989, ApJ 184, 271}

\noindent{Rotschild, R.E., Blanco, P.R., Gruber, D.E. et. al. 1998, ApJ, 496, 538}

\noindent{Scott, D.M, Finger, M.H., Wilson, R.B. et al. 1997, ApJ, 488, 831}

\noindent{Slettebak, A., 1988, PASP 100,770}

\noindent{Stella, L., White, N.E., Rosner, R., 1986, ApJ 308, 669}

\noindent{Valinia, A., Marshall, F.E., 1998, ApJ 505, 134}
 
\noindent{White, N.E, Swank, J.H., Holt, S.S. 1983, ApJ 270,711}
 
\noindent{Wilson, C.A., Finger, M.H., Coe, M.J., et al., 2002, ApJ 570, 
287}

\noindent{Zhang, Z.N, Yu, W., Zhang, W., 1998, ApJL 494, 71}
 
\end{document}